# INTERFACE TRAPS IN GRAPHENE FIELD EFFECT DEVICES: EXTRACTION METHODS AND INFLUENCE ON CHARACTERISTICS

G.I. Zebrev, E.V. Melnik, A.A. Tselykovskiy

Department of Micro- and Nanoelectronics, National Research Nuclear University MEPHI, 115409, Kashirskoe sh., 31, Moscow, Russia, gizebrev@mephi.ru

*Abstract*

We study impact of the near-interfacial oxide traps on the C-V and I-V characteristics of graphene gated structures. Methods of extraction of interface trap level density in graphene field-effect devices from the capacitance-voltage measurements are described and discussed. It has been found that the effects of electron-electron or hole-hole interactions and electron-hole puddles can be mixed in C-V characteristics putting obstacles in the way of uniquely determined extraction of the interface trap density in graphene. Influence of the interface traps on DC and AC capacitance and conductance characteristics of graphene field-effect structures is described. It has been shown that variety of widths of resistivity peaks in various samples could be explained by different interface trap capacitance values.

1. **Introduction**

Occurrence of charged defects in insulating layers nearby the conductive channels often referred to as "interface traps" or "border traps" is practically unavoidable generic problem of all field-effect devices [1, 2]. The charged oxide defects located near the graphene sheet cause scattering of carriers in the channel resulting in a decrease of free path time and carrier mobility due to the elastic Coulomb scattering mechanisms [3]; [4]. There exists another prominent effect of the traps in field-effect devices. It is well known that high density of interface traps suppress the electric field effect in gated structures and degrades the shapes of transfer I-V characteristics [1]. Interface traps also straightforwardly influence on the C-V gate characteristics due to fast carrier exchange between defects and channel during DC voltage ramp or AC small-signal input signal [5]. The extensive systematic experimental studies of the interface traps in graphene field-effect structures are still lacking. Therefore the role of fast interface traps in operation of graphene gated structure as a FET needs to be understood [6].

Rapid progress of graphene electronics has led to a growing need for simple and reliable methods of electric characterization of graphene field-effect structures. The problem is that many physical phenomena cannot manifest themselves directly, but only through the secondary device effects. By this reason the I-V curves of graphene and silicon FETs seem often to be very similar since their generic operation principle relies mainly upon classical electrostatic induction. Intrinsic fundamental parameters of devices are often reflected in experimental data as contaminated by many extrinsic occasional factors such as interface traps, parasitic resistance, capacitances, and technology-dependent defects. Without a systematic study of those factors, the validity of electrical data extracted from the FET characteristics cannot be considered accurate enough. Thus, data reported in the graphene literature should be taken with caution, as the majority of publications don't take any precautions to improve the quality of electrical data [7].

The chapter is organized as follows. Section 2 describes briefly a concept of interface traps as rechargeable oxide defects sensing the Fermi energy position in the channel. Section 3 is devoted to the model background equations where we examine the concepts of quantum capacitance, gate capacitance and channel capacitance. Interface trap level density extraction methods from C-V characteristics and modeling of the gate capacitance taking into account the effects of electron-electron interactions, electron-hole puddles and interface traps are described in Section 5. Influence of interface traps and ionizing irradiation on the graphene FET's characteristics is reviewed in Section 6.

## 2. Interface traps as near-interfacial rechargeable defects

Graphene sheet is located as a rule between the two insulated dielectrics. The interfaces between graphene and both top and bottom insulating layers are non-ideal and can have interface or near-interfacial defects which can significantly degrade device performance [8]. If the trapped charge constitutes a significant portion of the mobile charge in the graphene channel then recharging processes would significantly impede the change of the Fermi energy, degrading thusly transconductance and field-effect mobility in graphene transistors.

Near-interfacial traps (defects) are located exactly at the interface or in the underlying oxide typically within 1-3 nm from the interface. These defects can have generally different charge states and capable to be recharged by exchanging the carriers (electrons and holes) with the device channels. Due to the carrier exchange possibility the near-interfacial trap occupancy senses the Fermi level position in graphene (see Fig.1).

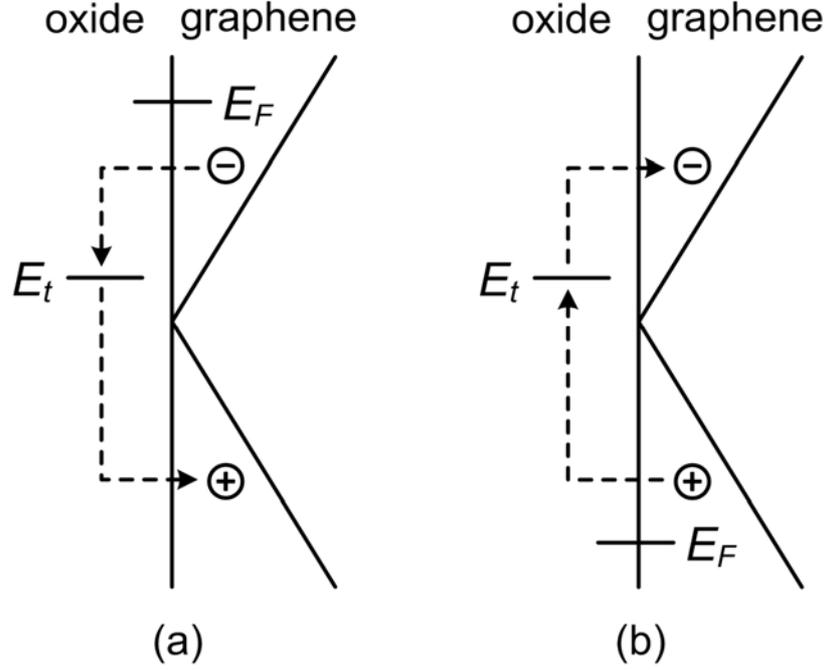

Fig. 1. Emptying and filling of the oxide traps located at or near the interface at different Fermi level positions in graphene.

Each gate voltage in graphene gated structure corresponds to the respective position of the Fermi level in graphene sheet with own equilibrium defect's occupation and quasi-equilibrium total trapped charge $Q_t(\varepsilon_F)$. The traps rapidly exchanging the carriers with the graphene are often referred as to the (fast) interface traps ($N_{it}$) [1, 5]. Interface trap capacitance per unit area $C_{it}$ (F/cm$^2$) and the interface trap level density $D_{it}(\varepsilon)$ (cm$^{-2}$ eV$^{-1}$) is defined as differential response of trapped charge to Fermi energy variation

$$C_{it}(\varepsilon_F) \equiv \frac{d}{d\varepsilon_F}\left(-Q_t(\varepsilon_F)\right) = e^2 D_{it}(\varepsilon_F). \tag{1}$$

Notice that quantum capacitance $C_Q = ed(n_e - n_h)/d\varepsilon_F$ is defined for mobile delocalized carriers in graphene layer. Typically (but not necessary) the trapped and mobile carriers are in quasi-equilibrium due to charge exchange between the traps and channel and have common Fermi energy. Due to equilibrium or non-equilibrium recharging processes the performance characteristics of graphene field-effect transistors are sensitive to the interface traps close to the graphene surface [9, 10]. It is useful to note that 1 fF/µm$^2 \cong 6.25 \times 10^{11}$ cm$^{-2}$ eV$^{-1}$.

## 3. General Background

### *3.1. Electrostatics of graphene single gated structure*

Description of planar electrostatics of a single-gate graphene structure is based on electric charge neutrality of the whole field-effect structure $N_G + N_t = n_e - n_h$, where $N_G$ is the area density of (positive) charge located on the gate, $N_t$ is the charged defect density (cm$^{-2}$) which is conditionally assumed to be positively charged ($Q_t = eN_t$), $n_S = n_e - n_h$ is the net charge density as imbalance of electron and hole concentrations in graphene.

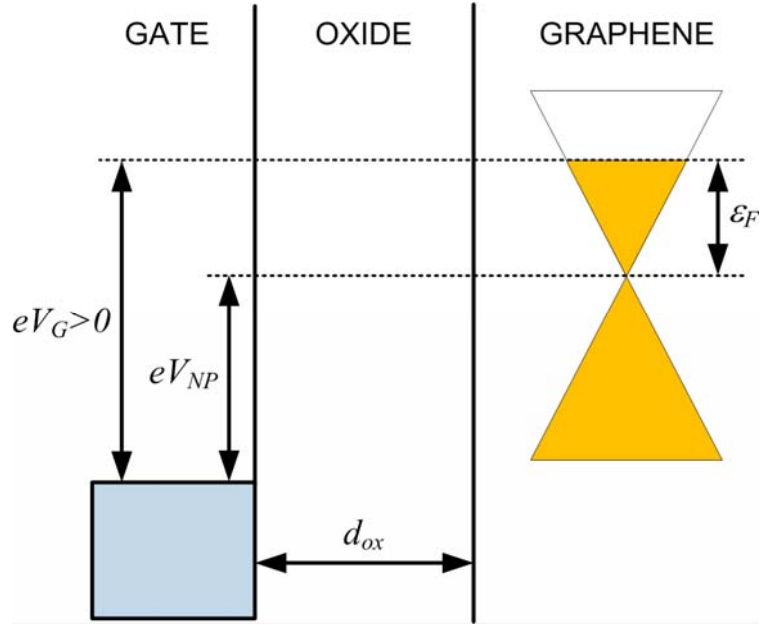

Fig. 2. Energy band diagram of graphene single-gated field-effect structure.

Fig. 2 shows a specific view of energy band diagram of the graphene gated structure. The basic equation of graphene planar electrostatics for single-gated structure can be written down in a form

$$e(V_G - V_{NP}) = \varepsilon_F + \frac{e^2 n_S}{C_{ox}} + \frac{e^2 \left(N_t(\varepsilon_F = 0) - N_t(\varepsilon_F)\right)}{C_{ox}}, \qquad (2)$$

where $C_{ox} = \varepsilon_{ox} \varepsilon_0 / d_{ox}$ is the capacitance (F/cm$^2$) of the insulating oxide with thickness $d_{ox}$ and dielectric constant $\varepsilon_{ox}$. The charge neutrality point (CNP) gate voltage (Dirac point) corresponding to the minimum of capacitance or conductivity $V_{NP}$ is determined by the work function difference between the gate material and graphene sheet $\varphi_{GG}$, and also by an equilibrium density of the charged near-interfacial defects at the CNP

$$V_{NP} = \varphi_{GG} - \frac{eN_t(\varepsilon_F = 0)}{C_{ox}}. \tag{3}$$

Taking for brevity without loss of generality $V_{NP} = 0$ and assuming constant density of the trap states one reads $e^2(N_t(\varepsilon_F = 0) - N_t(\varepsilon_F)) \cong C_{it}\varepsilon_F$. Taking into $n_S = \varepsilon_F^2/\pi\hbar^2 v_0^2$ ($v_0$ is the graphene Fermi speed) the basic equation of graphene planar electrostatics can be written down a in a form [11]

$$eV_G = \varepsilon_F + \frac{e^2 n_S}{C_{ox}} + \frac{C_{it}}{C_{ox}}\varepsilon_F \equiv m\varepsilon_F + \frac{\varepsilon_F^2}{2\varepsilon_a}, \tag{4}$$

where we have introduced for convenience a dimensionless "ideality factor"

$$m \equiv 1 + \frac{C_{it}}{C_{ox}}. \tag{5}$$

Every graphene field-effect structure is characterized by the unique energy

$$\varepsilon_a = \frac{C_{ox}}{C_Q}\varepsilon_F = \frac{\pi\hbar^2 v_0^2 C_{ox}}{2e^2} = \frac{\varepsilon_{ox}}{8\alpha_G}\frac{\hbar v_0}{d_{ox}}, \tag{6}$$

where the graphene "fine structure constant" is defined as ( in SI units)

$$\alpha_G = \frac{e^2}{4\pi\varepsilon_0 \hbar v_0}. \tag{7}$$

The characteristic energy $\varepsilon_a$ varies in the range from ~ $10^{-3}$ eV at $d_{ox}$ ~ 200 nm and $\varepsilon_{ox} = 4$ (SiO$_2$) to $\varepsilon_a$ ~ 0.5 eV at $d_{ox}$ ~2 nm and $\varepsilon_{ox} = 16$ (HfO$_2$).[11]

### 3.2. Quantum capacitance

Capacitance measurements provide important information about density of states of the localized and the mobile states at the Fermi energy in the 2D systems. The former ($C_{it}$) is associated with the interface traps which are capable to change their occupancy with gate bias changes and have energy levels distributed throughout the insulator bandgap. The latter ($C_Q = e^2 dn_S / d\varepsilon_F$) is connected with the thermodynamic compressibility and is often referred to as quantum capacitance. The concept of "quantum capacitance" was introduced in order to develop an equivalent circuit model for devices that incorporate a highly conducting two-dimensional (2D) electron gas [12].

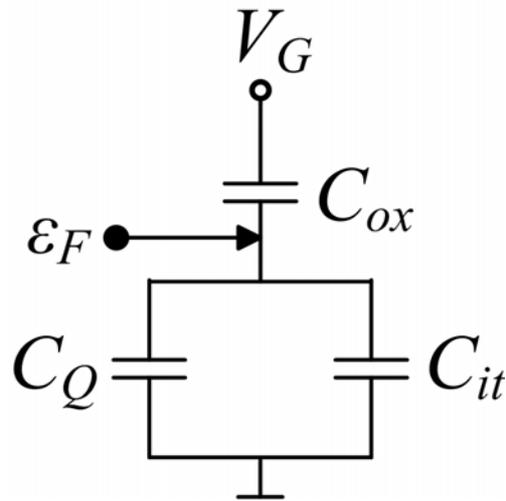

Fig. 3 Equivalent electric circuit of single-gated graphene device.

In contrast to conventional silicon MOS structures the $C_Q$ in zero bandgap graphene is always comparable with $C_{it}$ and cannot be neglected in electric equivalent circuit (see Fig.3). The quantum capacitance can be considered as a direct generalization of the "inversion layer capacitance"[13] in the silicon MOSFETs to the case of strictly one-subband filling. The inversion layer ("quantum") capacitance plays rather minor role in the silicon MOSFETs since it is negligibly low in the subthreshold operation mode (arising due to finite bandgap in the Si) and extremely high in the above threshold strong inversion regime. In the former case the quantum capacitance in MOSFETs is masked by the parasitic interface trap and the depletion layer capacitances connected in parallel in the equivalent electric circuit, and in the latter case it is insignificant due to the series connection with the gate insulator having typically lesser capacitances for high carrier densities in inversion layers.

Generally the total charge density in graphene consists of the electron $n_e$ and the hole $n_h$ components which are calculated exactly for ideal homogeneous graphene as function of the Fermi energy (chemical potential) $\varepsilon_F$

$$n_e \pm n_h = \frac{2}{\pi}\left(\frac{k_B T}{\hbar v_0}\right)^2 \left(\pm Li_2\left(-e^{\frac{\varepsilon_F}{k_B T}}\right) - Li_2\left(-e^{-\frac{\varepsilon_F}{k_B T}}\right)\right), \tag{8}$$

$Li_2(x)$ is the poly-logarithm function of the 2-nd order [14], $v_0$ (~$10^8$ cm/s) is the characteristic graphene speed. This relationship yields an exact value of residual intrinsic concentration $n_i = n_e(\varepsilon_F = 0) + n_h(\varepsilon_F = 0) = (\pi/3)(k_B T/\hbar v_0)^2$ in ideal graphene at the CNP. Quantum capacitance is calculated in ideal graphene as derivative of the net charge density in the channel with respect to chemical potential

$$C_Q = e^2 \frac{d(n_e - n_h)}{d\varepsilon_F} = \frac{2}{\pi}\left(\frac{e^2}{\hbar v_0}\right)\frac{k_B T}{\hbar v_0}\ln\left(2 + 2\cosh\left(\frac{\varepsilon_F}{k_B T}\right)\right) = C_{Q\min}\left(1 + \frac{\ln\cosh[\varepsilon_F/2k_B T]}{\ln 2}\right). \quad (9)$$

Note that the quantum capacitance $C_Q$ is an even function of the Fermi energy with the minimum value determined by the thermal de Broglie wavelength $\lambda_{dB} = k_B T/\hbar v_0$ in graphene

$$C_{Q\min} = \frac{2}{\pi}\frac{e^2}{\hbar v_0}\frac{k_B T}{\hbar v_0}\ln 4 \sim 10 \text{ fF/}\mu\text{m}^2. \quad (10)$$

For a relatively high doping case ($|\varepsilon_F| \gg k_B T$) we have an *approximate* relationship for quantum capacitance $C_Q \cong 2e^2\varepsilon_F/\pi\hbar^2 v_0^2$.

Explicit differentiation of total sum of electron and hole densities in Eq.8 leads to the an *exact* relation [11]

$$e^2\frac{d(n_e + n_h)}{d\varepsilon_F} = \frac{2e^2\varepsilon_F}{\pi\hbar^2 v_0^2}, \quad (11)$$

which is valid both for positive and negative Fermi energies at any finite temperature. Integrating Eq.11 one gets a simple *exact* relation for sum of carrier densities useful for conductivity calculation

$$N_S = n_e + n_h = \frac{\varepsilon_F^2}{\pi \hbar^2 v_0^2} + n_{res}. \qquad (12)$$

The integration constant (residual concentration at the Dirac point) $n_{res}$ turns out to be equal to the intrinsic concentration $n_i$ for ideal clean graphene. In this case the Eq.12 becomes an exact representation of the complicated form in Eq.8.

### 3.3. Quantum capacitance and electron-hole puddles

Electron-hole puddles in graphene are another consequence of presence of the near-interfacial charged defects in the underlying insulator [15]. Electron-hole puddles modify quantum capacitance and conductivity near the charge neutrality point increasing its minimum values. The observed minimums of small-signal C-V characteristics are also strongly influenced by the electron-hole puddles. The long-range potential fluctuation induced by charged near-interfacial defects distributed in uncorrelated way in the insulator can be described by Gaussian distribution function

$$P(u) = \frac{1}{\sqrt{2\pi \langle \delta u^2 \rangle}} \exp\left(-\frac{u^2}{2\langle \delta u^2 \rangle}\right), \qquad (13)$$

where $u$ is fluctuating potential reckoning from a mean value, $\langle \delta u^2 \rangle$ is the dispersion of potential fluctuation. The standard deviation for potential of uncorrelated near-interfacial defects can be assessed as [16]

$$\langle \delta u^2 \rangle = \frac{e^4}{\left(4\pi \varepsilon_0 \bar{\varepsilon}_{ox}\right)^2} \pi n_{imp} \qquad (14)$$

and generally to be determined by a sum of the positively and negatively charged defect densities $n_{imp} = n_{imp}^{(+)} + n_{imp}^{(-)}$; $\bar{\varepsilon}_{ox}$ is a half-sum of the dielectric constants for adjusted insulators. In the Thomas-Fermi approximation the local value of charge density can be written as

$$n_S = sgn(\varepsilon_F - u(\mathbf{r})) \frac{(\varepsilon_F - u(\mathbf{r}))^2}{\pi \hbar^2 v_0^2}, \qquad (15)$$

where $\varepsilon_F$ is a uniform equilibrium Fermi energy of the inhomogeneous channel.

At first we have to calculate the total net electric charge in graphene as function of $\varepsilon_F$ taking into account occurrence of the long-range potential fluctuation and electron-hole puddles:

$$\begin{aligned} Q^{(p)}(\varepsilon_F) &\cong \frac{e}{\pi \hbar^2 v_0^2} \left( \int_{-\infty}^{\varepsilon_F} (\varepsilon_F - u)^2 P(u) du - \int_{\varepsilon_F}^{\infty} (u - \varepsilon_F)^2 P(u) du \right) = \\ &= \frac{e}{\pi \hbar^2 v_0^2} \left[ \left( \langle \delta u^2 \rangle + \varepsilon_F^2 \right) erf\left( \frac{\varepsilon_F}{\sqrt{2 \langle \delta u^2 \rangle}} \right) + \varepsilon_F \sqrt{\langle \delta u^2 \rangle} \sqrt{\frac{2}{\pi}} \exp\left( -\frac{\varepsilon_F^2}{2 \langle \delta u^2 \rangle} \right) \right]. \end{aligned} \qquad (16)$$

Then the quantum capacitance accounting the electron-hole puddles becomes

$$C_Q^{(p)} = e \frac{\partial Q^{(p)}}{\partial \varepsilon_F} = \frac{2e^2 \varepsilon_F}{\pi \hbar^2 v_0^2} \left[ erf\left( \frac{\varepsilon_F}{\sqrt{2 \langle \delta u^2 \rangle}} \right) + \frac{\sqrt{\langle \delta u^2 \rangle}}{\varepsilon_F} \sqrt{\frac{2}{\pi}} \exp\left( -\frac{\varepsilon_F^2}{2 \langle \delta u^2 \rangle} \right) \right]. \qquad (17)$$

This equation differs from $C_Q = 2\varepsilon_F / \pi\hbar^2 v_0^2$ only in the vicinity of the CNP ($\varepsilon_F < |\langle \delta u^2 \rangle^{1/2}|$). It does not contain temperature since to be only valid for a condition $k_B T < \langle \delta u^2 \rangle^{1/2}$. At the CNP we have the minimum quantum capacitance in disordered graphene

$$C_{Q\min}^{(p)} = C_Q^{(p)}\left(\varepsilon_F = 0\right) = \frac{2e^2}{\pi\hbar^2 v_0^2} \sqrt{\frac{2\langle \delta u^2 \rangle}{\pi}}, \qquad (18)$$

instead of Eq.10 which is valid for $k_B T > \langle \delta u^2 \rangle^{1/2}$.

### *3.4. Gate capacitance*

Capacitance-voltage measurements are very important in providing information about the gated field-effect structures. The shapes both of C-V and I-V characteristics are determined by the dependence of the Fermi energy position in graphene on the gate voltage. Taking derivative of Eq. 2 with respect to Fermi energy, we have

$$\frac{dV_G}{d\varepsilon_F} = 1 + \frac{C_Q + C_{it}}{C_{ox}}. \qquad (18)$$

Basic measurable small-signal parameter of the field-effect structures is the differential gate capacitance. The frequency-dependent input gate capacitance is determined as derivative of the gate charge density with respect to $V_G$:

$$C_G = e\left(\frac{\partial N_G}{\partial V_G}\right) = e\frac{dN_G/d\varepsilon_F}{dV_G/d\varepsilon_F} = \frac{C_{it}(\omega)+C_Q}{1+\frac{C_Q+C_{it}}{C_{ox}}} = \left(\frac{1}{C_{ox}}+\frac{1}{C_Q+C_{it}(\omega)}\right)^{-1}. \quad (19)$$

In contrast to generally time-dependent interface trap response the quantum capacitance is assumed to be instantaneous, i.e. the frequency-independent even in high frequency applications. Notice that $C_G$ is an increasing function of $C_{it}$ and decreases with the gate small-signal frequency increase since the $C_{it}$ logarithmically diminishes for high a.c. frequency due to carrier exchange rate suppression under fast gate signal. The carrier exchange between gapless graphene monolayer and near-interfacial traps in the gate dielectric occurs mainly due to elastic tunneling at the Fermi level (at low temperatures) or thermally activated tunneling at room temperatures. The full response of the traps is determined as superposition of responses of single traps. This superposition of the Lorentzian peaks with exponentially wide range of the recharging time results in a logarithmic dependence on AC frequency [17] [18]

$$C_{it}(\omega) = C_{it}\left(1-(\lambda/l)\ln\omega\tau_r\right), \quad (21)$$

where $C_{it}$ is the low-frequency interface trap capacitance, $l$ is a thickness of the trap location (assumed to be distributed uniformly, typically < 3 nm), $\lambda$ is the tunneling length or inverse imaginary wavevector (~ 0.1 nm). This means that a lower portion of the interface traps remains active at high frequencies [19, 20, 21].

Additional problems arise when modeling characteristics at different temperatures. Typically the characteristic carrier exchange times decrease at elevated measurement temperatures leading to an increase in the gate capacitance with temperature [21, 22, 23]. It has been found that the gate capacitance normally increases as temperature increases also due to appreciable increase of dielectric constant of the gate insulator [22].

Generally the gate capacitance decreases with interface trap density increase for all gate voltages. Loosely speaking, the capacitance of the two-plate condenser is determined by the effective length of the electric field lines between the two capacitor plates. In contrast to conventional metallic plates, the field line can penetrate through graphene sheet decreasing thereby capacitance of the graphene field-effect structures. This occurs due to moderate values of quantum capacitance in graphene especially near charge neutrality point. Mean values of field penetration beyond the graphene sheet are of order of distance between the carriers in graphene $n_S^{-1/2} \propto C_Q^{-1}$. Interface traps suppress the field line penetration beyond the graphene plate making the capacitor more like a conventional metallic plate case.

### 3.5. Channel capacitance

We discriminate distinctly the gate capacitance and "the channel capacitance" [20] (see also [24]) which is defined as the derivative of total carrier density $N_S = n_e + n_h$ in the channel with respect to the gate voltage:

$$C_{CH} = e\left(\frac{\partial N_S}{\partial V_G}\right) = e\frac{dN_S/d\varepsilon_F}{dV_G/d\varepsilon_F} = \frac{C_{ox}\varepsilon_F/\varepsilon_a}{1+(C_Q+C_{it})/C_{ox}}. \qquad (22)$$

Strictly speaking, $C_{CH}$ is not the electric capacitance and cannot be represented in a form of an equivalent electric circuit. In contrast to the gate capacitance which corresponds to the small-signal capacitance characteristics; the channel capacitance determines the small-signal current parameters. For example, for linear regime when drain current of FET $I_D = Wen_S v_{dr}$ is expressed through the carrier's drift velocity $v_{dr}$, the transconductance reads as follows

$$g_m = \frac{\partial I_D}{\partial V_G} = W C_{CH} v_{dr}. \qquad (23)$$

The gate capacitance is an even and non-zero function of the gate voltage at any finite temperature while the $C_{CH}$ is the odd function changing the sign at the charge neutrality point. Beyond the vicinity of the CNP we have $C_Q = C_{ox}(\varepsilon_F/\varepsilon_a)$, and the gate and the channel capacitances turn out to be connected in graphene gated structures through the relation

$$\frac{C_G}{C_{CH}} = 1 + \frac{C_{it}}{C_Q}. \qquad (24)$$

All relationships for the differential capacitances remain valid for any form of the interface trap energy spectra. In an ideal case, the channel capacitance $C_{CH}(V_G)$ should be symmetric with refer to the neutrality point implying approximately flat energy density spectrum of interface traps. Generally, the channel capacitance is a more appropriate concept for I-V characteristic description whereas the gate capacitance is a directly measured quantity in the C-V measurements.

### 3.6. Fermi energy and charge density as functions of gate voltage

Solving algebraic Eq.4 one arrives at the relationship connecting explicitly the Fermi energy in graphene and gate voltage [11] with the gate insulator $\varepsilon_a$ (Eq.6) and the interface trap $m$ (Eq.5)

parameters:

$$\varepsilon_F = \left(m^2\varepsilon_a^2 + 2\varepsilon_a e|V_{GS} - V_{NP}|\right)^{1/2} - m\varepsilon_a. \qquad (25)$$

Fig. 4 shows the typical dependencies of the Fermi energy versus gate voltage at different interface trap capacitances.

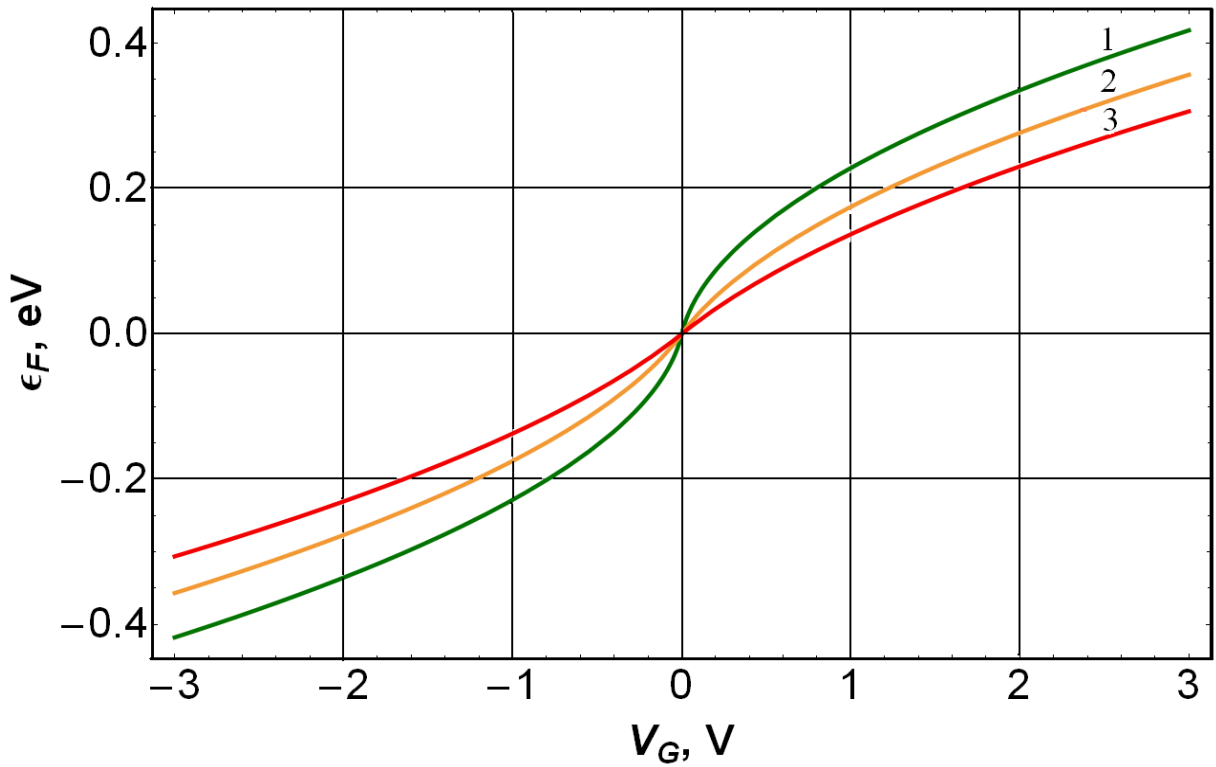

Fig. 4. Fermi energy in graphene simulated as functions of the gate voltage at different interface trap capacitances: (1)0; (2)10 $fF/\mu m^2$; (3) 20 $fF/\mu m^2$.

Combining Eqs. 2 and 25 the explicit relation for graphene charge density dependence on gate voltage can be written as follows

$$\frac{e^2 n_S}{C_{ox}} = eV_G - m\varepsilon_F = eV_G + m^2\varepsilon_a - m\left(m^2\varepsilon_a^2 + 2\varepsilon_a eV_G\right)^{1/2}. \tag{26}$$

Defining for brevity the characteristic voltage

$$eV_0 \equiv m^2\varepsilon_a = \left(1 + \frac{C_{it}}{C_{ox}}\right)^2 \varepsilon_a, \tag{27}$$

the latter equation can be rewritten as [25]

$$en_S(V_G) = C_{ox}\left(|V_G - V_{NP}| + V_0\left(1 - \left(1 + 2\frac{|V_G - V_{NP}|}{V_0}\right)^{1/2}\right)\right), \tag{28}$$

where the interface trap capacitance is taken into account [26].

### 3.7. Interactions and quantum capacitance

In the absence of interaction the quantum capacitance (or, compressibility) is proportional to the thermodynamic density of states and, consequently, to the inverse single electron spacing in the system. Therefore the compressibility and quantum capacitance are expected to decrease due to repulsive electron-electron or hole-hole interactions since it costs more energy to add particles into the system. Quantum capacitance depends on the Fermi velocity in graphene $v_0$. Therefore, when interactions are taken into account, it has been predicted that the linear dispersion law under the single-particle picture will change and the Fermi velocity will increase dramatically towards the Dirac point [27]. Thereby, the interactions reduce the density of states at the Dirac point, making the system less metallic. More specific, the interactions are generally expected to give rise to a logarithmic renormalization of the Fermi velocity ($n_{max} \sim 10^{15}$ cm$^{-2}$) [28]

$$v_0(n_S) = v_{00}\left[1 + \left(\frac{e^2}{4\pi\varepsilon_0 \hbar v_{00}}\right)\frac{\ln(n_{max}/n_S)}{8\varepsilon_{ox}}\right]. \tag{29}$$

Renormalization of the graphene Fermi velocity is the very important issue since $v_0$ is almost the only fundamental parameter in the theory entering into practically all observable quantities. The fact that it can vary as a function of energy should be taken into account for the correct interpretation of the experimental results [29].

4. **Interface traps influence on C-V characteristics**

   *4.1. Simple method of direct extraction of interface trap capacitance from C-V characteristics*

Influence of interface traps on C-V curves in field-effect devices is two-fold. Firstly, the total gate capacitance at a given Fermi energy in graphene increases with the interface trap density (AC response). Secondly, the gate voltage dependence on the Fermi energy $V_G(\varepsilon_F)$ is influenced by the interface trap recharging during the gate voltage ramp that leads to the stretch-out of the C-V curves along the gate voltage axis (DC response). All information about the energy level distribution of the interface trap density is contained in the dependence $V_G(\varepsilon_F)$. Knowing the quantum capacitance $C_Q$ and using Eq.19 the differential interface trap density as function of Fermi energy reads immediately as

$$C_{it}(\varepsilon) = C_{ox}\left(\frac{edV_G}{d\varepsilon_F} - 1\right) - C_Q(\varepsilon). \tag{30}$$

Capacitance-voltage measurements provide information about the Fermi energy as function of the gate voltage. Combining Eqs.18 and 19 one gets

$$\frac{d\varepsilon_F}{edV_G} = 1 - \frac{C_G}{C_{ox}}. \tag{31}$$

Using Eq.31 the Fermi energy at any applied gate voltage could be determined from integration of the experimental C-V curve (Berglund method [1, 5, 30])

$$\varepsilon_F(V_G) = \int_{V_{NP}}^{V_G} \left(1 - \frac{C_G(V_G')}{C_{ox}}\right) edV_G', \tag{32}$$

where the integration constant is chosen to be zero in graphene since the charge neutrality point at the capacitance minimum voltage $V_{NP}$ corresponds to the zero Fermi energy. This method is illustrated in Fig.5 where the C-V data from [31] are used as an example.

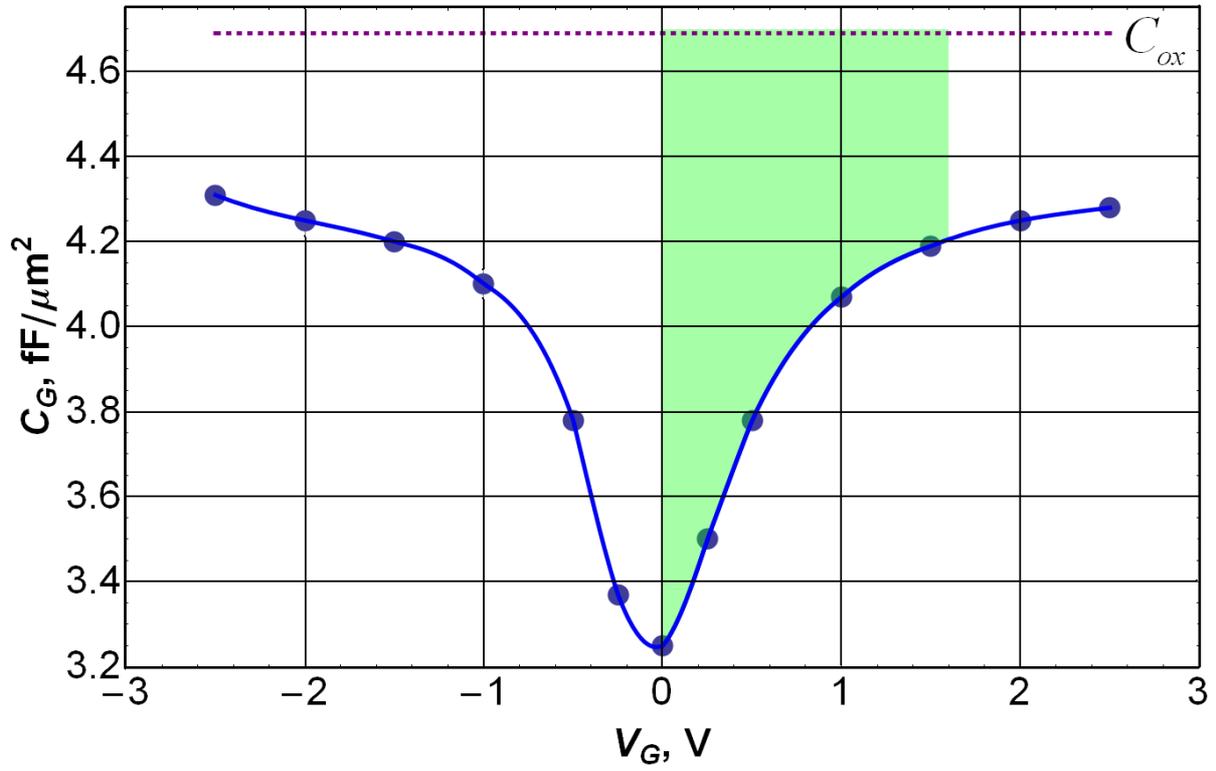

Fig. 5. Shadowed area corresponds to the Fermi energy as function of the gate capacitance $C_G(V_G)$. C-V data are taken from [31]($d_{ox}$ = 10 nm, $\varepsilon_{ox}$ = 5.3(Al$_2$O$_3$)).

Numerical analysis of the experimental C-V curve with the Eq.32 enables obtaining of dependencies $V_G(\varepsilon_F)$ and $dV_G/d\varepsilon_F$ (see Fig. 6) containing theoretically all information about interface trap spectrum.

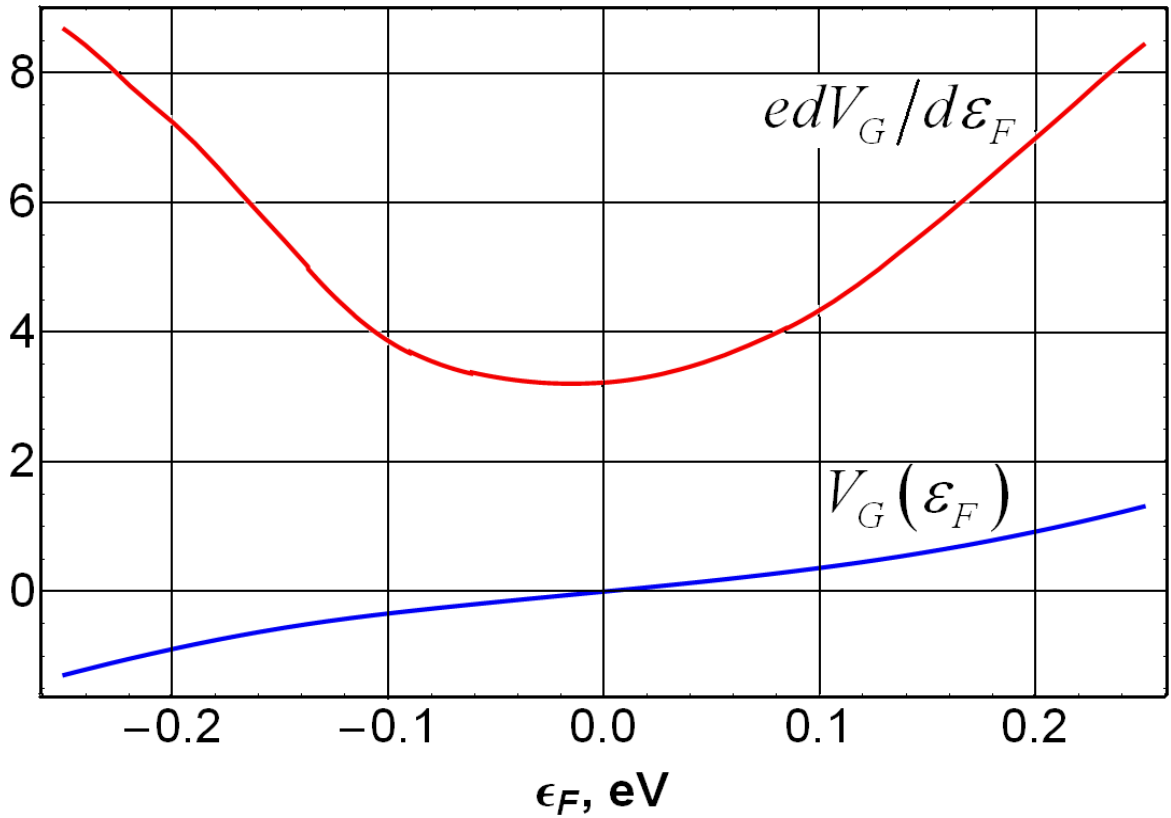

Fig.6 Dependencies $V_G(\varepsilon_F)$ (in Volts) and $edV_G/d\varepsilon_F$ (dimensionless) extracted numerically from the C-V data [31].

Sum of the quantum and interface trap capacitances as function of gate voltage can be obtained immediately from C-V characteristics

$$C_Q + C_{it} = C_G\left(\frac{1}{C_G} - \frac{1}{C_{ox}}\right)^{-1} = C_G \frac{edV_G}{d\varepsilon_F}. \qquad (33)$$

The dependence $V_G(\varepsilon_F)$ allows to calibrate the abscissa axes and to obtain $C_Q + C_{it}$ as function of the Fermi energy. Fig.7 shows comparison of the theoretically calculated dependence of quantum capacitance $C_Q(\varepsilon_F)$ and experimentally extracted sum of the quantum and the interface trap capacitances

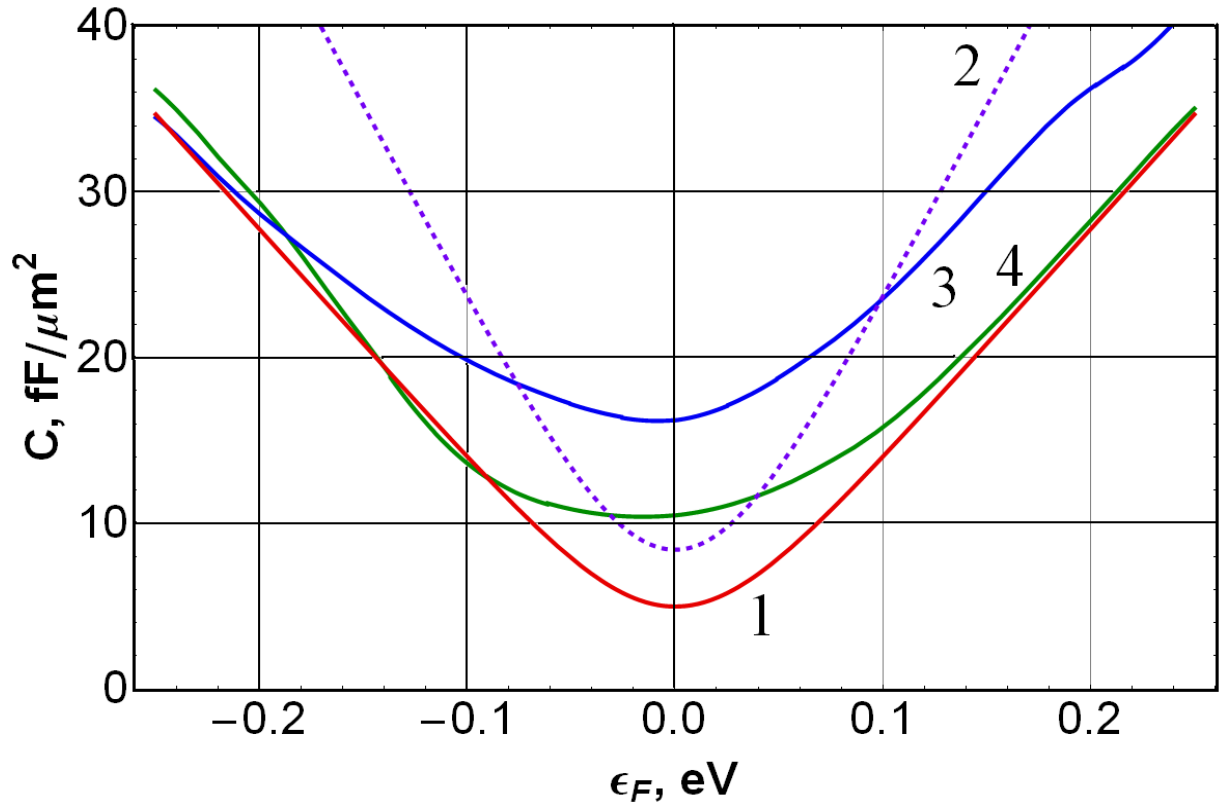

Fig.7 Comparison of quantum capacitance vs Fermi energy $C_Q(\varepsilon_F)$ calculated with $v_0 = 1.3\times10^8$ cm/s (curve 1), $v_0 =1.0\times10^8$ cm/s (curve 2, dashed line) and extracted curves $C_G(dV_G/d\varepsilon_F) = C_Q + C_{it}$ obtained numerically using Eq. 32 for experimental data taken from [32] (curve 3), and [31] (curve 4).

The separation results for $C_Q$ and $C_{it}$ are strongly dependent on *a priori* value of $v_0$. As can be seen in Fig.7, a model-independently extracted curve $C_G(dV_G/d\varepsilon_F)$ imposes limitation on numerical value of $v_0$ which assumed here to be constant. In particular the quantum capacitance calculated with $v_0=1.0 \times 10^8$ cm/s in an unphysical way exceeds $(C_Q + C_{it})_{exp}$ extracted from the experiments. A value $v_0 = 1.3 \times 10^8$ cm/s seems to be more appropriate for self-consistent description of the experimental C-V data. This circumstance may be explained by influence of electron-electron interaction which can significantly increase the effective Fermi velocity near the CNP.

Unfortunately as well as for the Si-MOSFET case the methods of absolute differential spectra extraction are very sensitive to experimental errors and occasional uncertainties in parameters and data such as asymmetry of C-V curves, uncertainty in dielectric properties of insulators etc. For these reasons the extraction of difference $\Delta D_{it}(\varepsilon)$, for example, before and after electric or irradiation stress could give more reliable results.

Another useful approach is to extract a restricted set of parameters in the frame of a simplified model. In practice one can utilize an effective $C_{it}$ parameter approximated as the mean of differential $C_{it}$ spectrum over all the range of gate voltage. Interactions may enter in this approach through an effective renormalized value of the Fermi velocity. For example, one can extract a single "non-ideality factor" $m$ and $v_0$ comparing the experimental dependence $\varepsilon_F(V_G)$ derived by the method described in this section and Eqs. 4 or 25. Figure 8 shows an example of such comparison with the fitted $C_{it}$ and $v_0$ for experimental C-V curve taken from [31]. Fitted parameters used are presented in Table 1.

Table 1. Fitted parameters

|  | $C_{it}$, fF/μm$^2$ | $C_{ox}$, fF/μm$^2$ | $v_0$, $10^8$ cm/s |
|---|---|---|---|
| [31] | 4.5 | 5.6 [31] | 1.35 |
| [32] | 10.0 | 4.7 [32] | 1.35 |

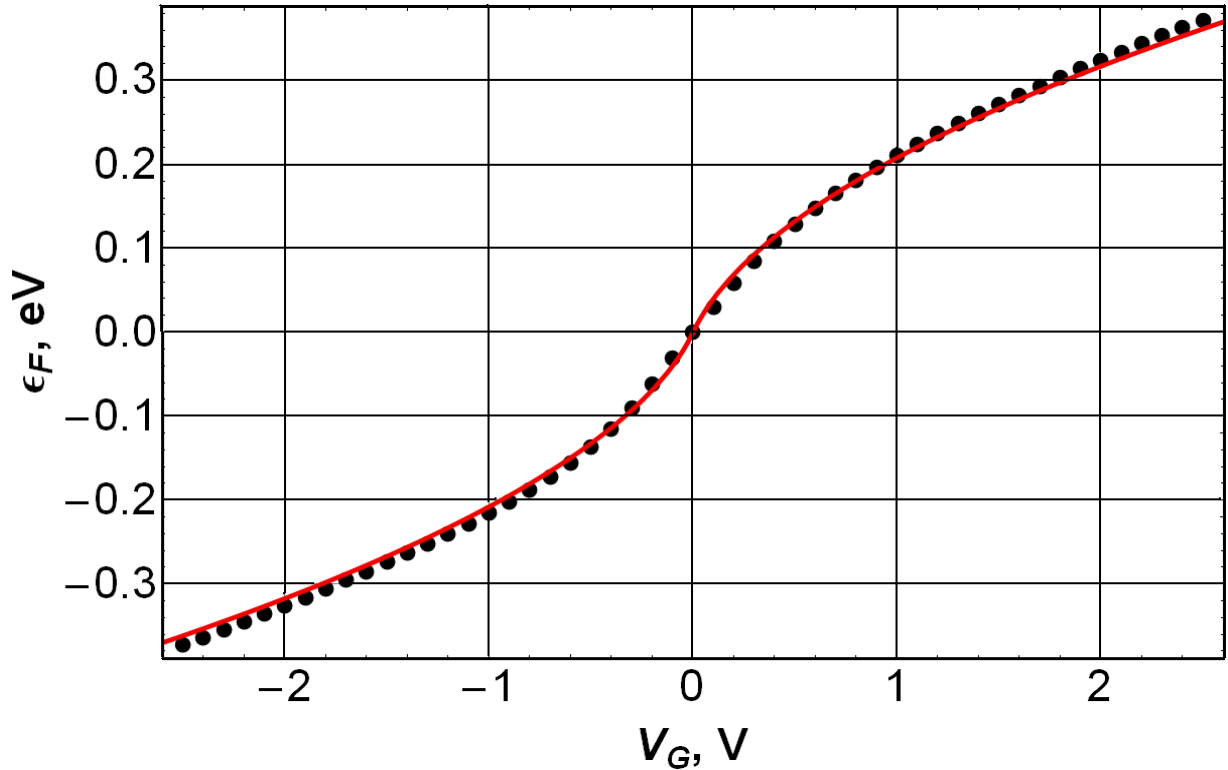

Fig.8 Comparison of experimental data extracted from [31] (points) with the simulated curve calculated with the fitted parameters presented in Table 1.

Notice that the experimental curve cannot be fitted at any $C_{it}$ value (including zero) for often used $v_0 = 1.0 \times 10^8$ cm/s. This is obvious evidence in favor of importance of interactions significantly increasing the Fermi velocity near the CNP.

### *4.2. Modeling of C-V characteristics taking into account interface traps, electron-hole puddles and interaction effects*

The gate capacitance curves $C_G(V_G)$ are formed by interplay of several poorly determined parameters having just as fundamental so and occasional character: the Fermi velocity renormalization constants $v_{00}$ and $n_{res}$, the interface trap capacitance $C_{it}$, and the typical dispersion of potential fluctuation in graphene $\langle \delta u^2 \rangle$. The problem of parameter extraction turns out to be obviously over-determined and does not allow a unique solution based at least on a single C-V curve analysis. Let us illustrate this by comparison of simulation results and the two set of experimental data in Fig.9 [31] and in Fig.10 [32]. Simulations were performed with three different sets of fitted parameters (see Tables 2, 3) corresponding to (a) taking into account influence simultaneously of velocity renormalization, electron-hole puddles and interface traps; (b) only with interactions and interface traps; (c) only with interface traps ignoring interactions and electron-hole puddles.

Table 2. Fitted parameters [31]

|   | $C_{it}$, fF/μm² | $D_{it}$, $10^{12}$ cm⁻² eV⁻¹ | $C_{ox}$, fF/μm² | $n_{res}$, $10^{11}$ cm⁻² | $v_{00}|v_0$(at CNP), $10^8$ cm⁻² | $\langle \delta u^2 \rangle^{1/2}$ meV |
|---|---|---|---|---|---|---|
| a | 4.5 | 2.8 | 5.6[31] | 3.5 | 0.59\|1.38 | 70 |
| b | 6.7 | 4.2 | 5.6[31] | 3.5 | 0.59\|1.38 | - |
| c | 4.5 | 2.8 | 5.6[31] | - | 1.15\|1.15 | - |

Table 3 Fitted parameters [32]

|   | $C_{it}$, fF/μm² | $D_{it}$, $10^{12}$ cm⁻² eV⁻¹ | $C_{ox}$, fF/μm² | $n_{res}$, $10^{11}$ cm⁻² | $v_{00}|v_0$(at CNP), $10^8$ cm⁻² | $\langle \delta u^2 \rangle^{1/2}$, meV |
|---|---|---|---|---|---|---|
| a | 6.15 | 3.8 | 4.7[32] | 7.3 | 0.86\|1.6 | 100 |
| b | 10.0 | 6.2 | 4.7[32] | 7.3 | 0.86\|1.6 | - |
| c | 10.0 | 6.2 | 4.7[32] | - | 1.45\|1.45 | - |

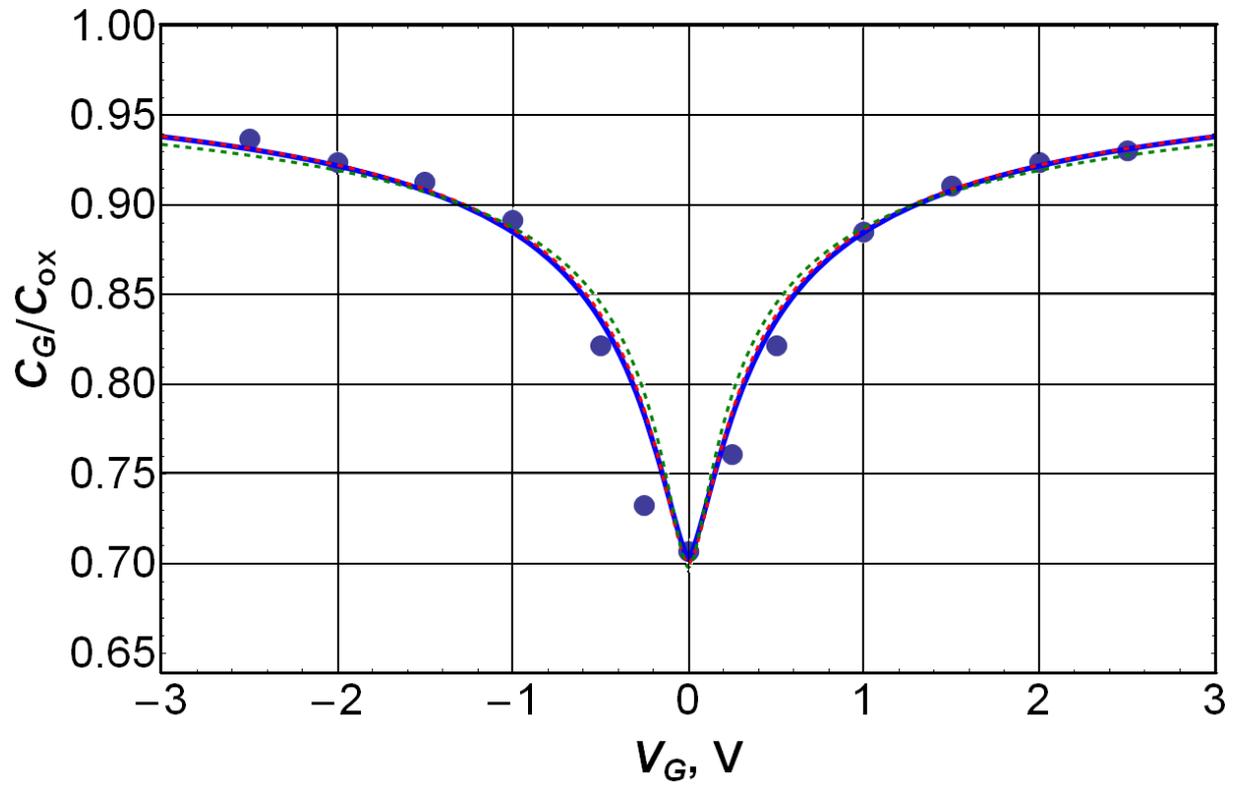

Fig.9. Simulated gate capacitance dependence in comparison with experimental points [31] with fitted parameters presented in Table 2: (a) solid line, (b) red and (c) green dashed lines.

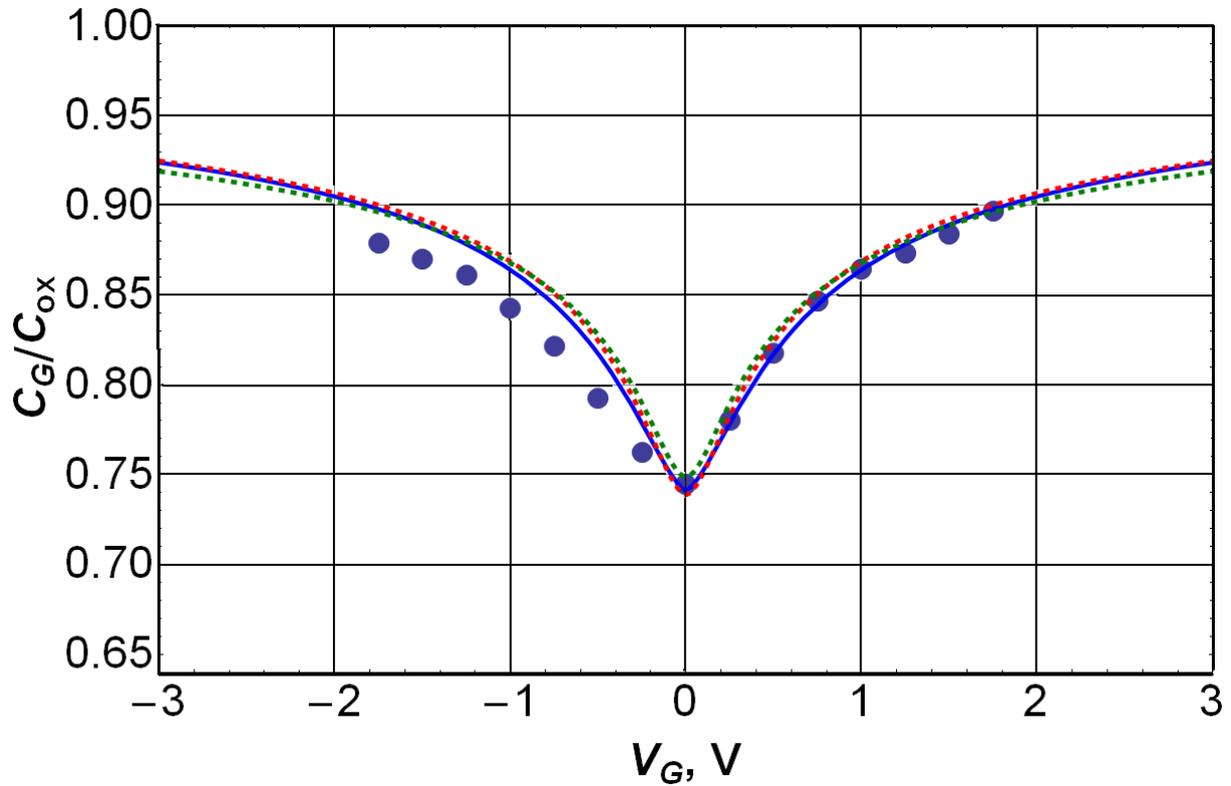

Fig.10. Simulated gate capacitance dependence in comparison with experimental points [32] with fitted parameters presented in Table 3: (a) solid line; (b) red and (c) green dashed lines.

As can be seen from Figs. 9, 10 the experimental points can be well fitted for all the three sets of fitted parameters with reasonable values. The simulation results are rather sensitive to all of three influence mechanisms and at the same time these mechanisms can easily compensate each others. Electron-hole puddles have a significant impact on the C-V curves of graphene gated structures only near the neutrality point. As well as the interface traps, the electron-hole puddles tend to increase the gate capacitance at the neutrality point, mixing the effects and making it difficult to separate them experimentally. Particularly, the neglect of the electron-hole puddle contribution near the CNP can be compensated by $C_{it}$ enhancement. An increase in $v_0$ at the

CNP due to interactions would lead to a decrease in $C_G$ ($V_{NP}$) and can be compensated by enhancement of $C_{it}$ and/or electron-hole puddle contribution. Branches of the C-V curves are practically insensitive to the interface traps and electron-hole puddles far away from the CNP but are sensitive to the interaction effects.

## 5. Interface traps and conductivity

The effect of traps on the conduction and mobility of single-layer graphene is briefly discussed in this section.

### 5.1. Low field conductivity

Assuming the zero contact resistance, equal mobility both for electrons and holes and taking into account Eq. 12 the low-field conductivity can be written as

$$\sigma_0 = e\mu_0 (n_e + n_h) = e\mu_0 \left( \frac{\varepsilon_F^2}{\pi \hbar^2 v_0^2} + n_{res} \right), \tag{34}$$

where $n_{res}$ is the residual carrier concentration at the CNP which assumed to be exactly equal to the intrinsic concentration $n_i$ in ideally homogeneous graphene or the total sum of carrier concentration in electron-hole puddles at the CNP. Using Eq.25 for $\varepsilon_F(V_G)$ one obtains

$$\sigma_0 = e\mu_0 n_{res} + \mu_0 C_{ox} \left( |V_G - V_{NP}| + V_0 - V_0 \left[ 1 + \frac{2|V_G - V_{NP}|}{V_0} \right]^{1/2} \right), \tag{35}$$

where conductivity as function of the gate voltage turns out to be explicitly dependent through the characteristic voltage $V_0$ both on the oxide parameter $\varepsilon_a$ and on the interface trap capacitance $C_{it}$. To obtain the conductivity of inhomogeneous graphene channel we have to calculate the mean sum of electron and hole concentrations in presence of long-range potential fluctuations

$$N_S^{(p)}(\varepsilon_F) \cong \frac{e}{\pi\hbar^2 v_0^2}\left(\int_{-\infty}^{\infty}(\varepsilon_F - u)^2 P(u)du\right) = \frac{\langle \delta u^2 \rangle + \varepsilon_F^2}{\pi\hbar^2 v_0^2}. \qquad (36)$$

Obviously, the Eq.36 implies that the residual carrier concentration at the CNP in Eq.35 [4] is determined immediately by the potential fluctuation dispersion

$$n_{res} = N_S^{(p)}(\varepsilon_F = 0) = \frac{\langle \delta u^2 \rangle}{\pi\hbar^2 v_0^2}. \qquad (37)$$

### 5.2. Modeling of graphene low-field resistivity

We have examined for illustration the experimental results presented in Ref. [33] where the resistivity $\rho = 1/e\mu_0 N_S$ was measured as function of gate voltage for different graphene samples. Resistivity for different samples was simulated using analytic Eq. 35 with several fitted parameters. We have simulated these results using experimental mobilities $\mu_0$ extracted by the authors of this paper. Excepting trivial $V_{NP}$, we fit only the interface trap capacitance $C_{it}$ and the residual concentration $n_{res}$ for three different samples. The total charged defect density $n_{imp}$ then has been recalculated using Eqs. 37 and 14. Comparison of experimental and simulated results is shown in Fig.11.

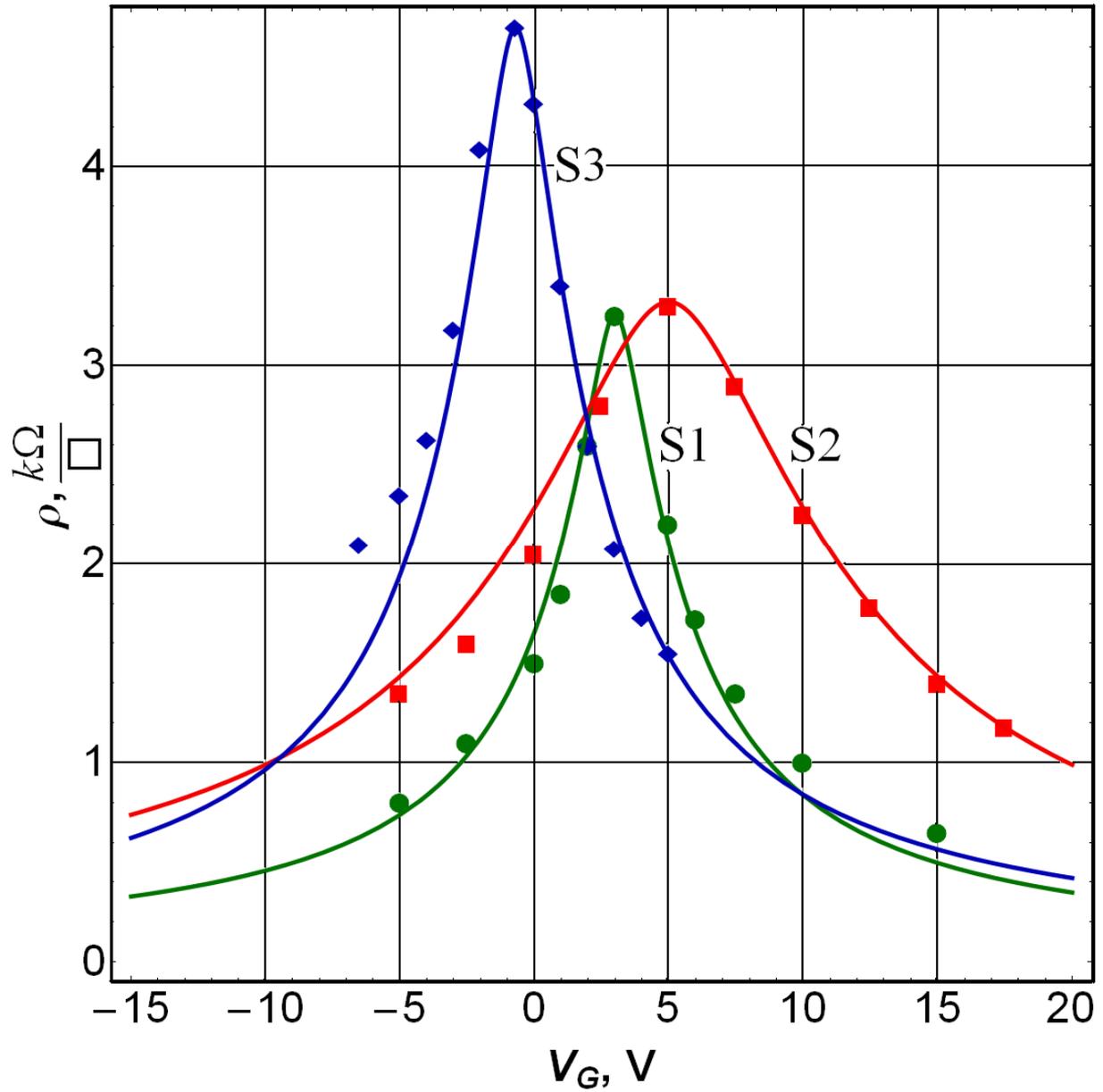

Fig. 11 Comparison of experimental resistivity [33] (points) as functions of gate voltage for different samples S1, S2, S3 [33] and simulated curves calculated using Eq.35 with fitted parameters presented in Table 4. Carrier mobilities were taken from [33].

The extracted parameters for different samples are summarized in Table 4.

Table 4. Fitted parameters

| sample | $C_{it}$, fF/μm$^2$ | $n_{res}$, $10^{11}$ cm$^{-2}$ | $n_{imp}$, $10^{11}$ cm$^{-2}$ | $\mu_0$, cm$^2$/Vs |
|---|---|---|---|---|
| S1 | 3.6 | 0.67 | 1.5 | 17500 [33] |
| S2 | 9.5 | 1.6 | 3.5 | 9300 [33] |
| S3 | 4.4 | 0.63 | 1.4 | 12500 [33] |

Simulation results exhibit an excellent agreement with the experiment in description of resistivity near the vicinity of the "Dirac peak" at reasonable values of parameters. This suggests that the widths of Dirac peaks are determined mainly by the interface trap density. Behavior of the resistivity dependencies at large $|V_G - V_{NP}|$ (where $\rho \leq 1$ kΩ/sq) is typically influenced by the contacts. Unfortunately the typical values of state-of-the-art graphene-metal contacts may be as high as hundreds of Ohm×μm and larger [34].

## 6. Influence of interface traps on small signal characteristics of graphene FETs

### 6.1. Transconductance and field-effect mobility

In contrast to the "true" low-field mobility $\mu_0$ depending only on microscopic scattering mechanisms, the transconductance and the field-effect mobility should be considered as device characteristics depending as well on the gated structure parameters. For example, the transconductance is defined as $g_m = \partial I_D / \partial V_G$, and in most cases has been extracted from the slope of the direct drain current dependence on the gate voltage [35, 36, 37, 38]. For the low-field linear regime the transconductance ($I_D = (W/L)\mu_0 e N_S V_D$) is determined by channel capacitance $C_{CH}$. Far away from the CNP we have the relationship [39]

$$g_m = e\frac{W}{L}\mu_0 V_D C_{CH} = e\frac{W}{L}\mu_0 V_D \frac{C_Q}{1+\frac{C_{it}+C_Q}{C_{ox}}} = e\frac{W}{L}\mu_0 V_D \frac{C_{ox}}{1+\frac{C_{it}+C_{ox}}{C_Q}}, \qquad (38)$$

which shows that the transconductance as well as the $C_{CH}$ decreases with increasing of interface trap capacitance. Eq.38 describes "concave" portion of I-V transfer characteristics until the curve becomes convex i.e. tending to saturation due to extrinsic factors (e.g. contact resistance). All field-effect devices are prone to the interface trap instabilities under influence of electric stresses or exposure of ionizing irradiation. The transconductance at fixed mobility $\mu_0$ has an important property following immediately from Eq.38. Any alteration of the interface trap capacitance under external impact $C_{it} \to C_{it} + \Delta C_{it}$ leads to a renormalization of the transconductance

$$g_m(C_{it}+\Delta C_{it}) = \frac{g_m(C_{it})}{1+\frac{\Delta C_{it}}{C_Q+C_{ox}+C_{it}}}. \qquad (39)$$

The greater an initial interface trap density $C_{it}$, the lesser impact of the added $\Delta C_{it}$. Transconductance is closely connected to the so-called field-effect mobility $\mu_{FE}$ normally defined as [40]

$$\mu_{FE} = \frac{L}{W}\frac{g_m}{C_{ox}V_D}. \qquad (40)$$

Taking into account Eq.38 this implies

$$\mu_{FE} = \mu_0 \frac{C_{CH}}{C_{ox}} = \frac{\mu_0}{1 + \frac{C_{it} + C_{ox}}{C_Q}}. \tag{41}$$

Eq.41 connects the field-effect mobility $\mu_{FE}$ depending particularly on charge exchange with the extrinsic traps (defects in the gate oxides, chemical dopants etc.) and mobility $\mu_0$ depending only on "microscopic" scattering mechanisms. Thus the simple $g_m$ method is not appropriate for true mobility extraction in real structures, and $\mu_0$ tends to be underestimated due to simultaneous recharging of the interface traps. Once the traps distribution could be determined, it's the true mobility can be extracted out and a more accurate characterization of electric transport in graphene can be achieved [9].

Due to frequency-dependent interface trap response, the traditional DC measurements are not sufficient when considering graphene transistor for high frequency (> GHz) circuit design in RF applications. It has been found that RF I-V curves show 50% increase in transconductance as compared to their DC I-V characteristics [19]. The increase in transconductance is attributed to reduced $C_{it}$ at GHz frequencies [19, 20]. Generally, the field-effect mobility is an increasing function of charge density with the zero minimum at the CNP due to an increase in quantum capacitance. Field-effect mobility usually never reaches its maximum value because of impact of extrinsic factors such as the parasitic contact resistances.

### *6.2. Cutoff frequency and logarithmic swing*

The cut-off frequency $f_T$ defined as the frequency at which the gain becomes unity is the most widely used figure-of-merit for RF devices and is, in effect, the highest frequency at which a FET is useful in RF applications. Omitting for brevity the problem of parasitic capacitances and series resistances in the source-drain circuit, the cut-off frequency $f_T$, i.e. the frequency at which the short-circuit current gain becomes unity can be written as [38]

$$f_T = \frac{g_m}{2\pi C_{GG}}, \qquad (42)$$

where $C_{GG} = C_{GS} + C_{GD}$, $C_{GS}$ and $C_{GD}$ are the gate-source and the gate-drain capacitances in equivalent capacitance circuit. Notice that $C_{GD} \ll C_{GS}$ and $C_{GS} \leq C_G$ for typical applications. Recalling that $g_m \propto C_{CH}$ one gets [20]

$$f_T \propto \frac{g_m}{C_G} \propto \frac{C_{CH}}{C_G} \propto \frac{1}{1+\frac{C_{it}}{C_Q}}. \qquad (43)$$

As can be seen from Eq.43, the cutoff frequency is a decreasing function of $C_{it}$, although an impact of $C_{it}$ diminishes at large charge density in the graphene channel.

Another important FET parameter is the logarithmic swing $S$ which characterizes $I_{ON}/I_{OFF}$ ratio and capability to modulate transistor's conductance [1]. The logarithmic swing (in Volts per drain current decade) equals numerically to the gate voltage alteration needed for current change by an order

$$SS \equiv \left(\frac{d(\log_{10} I_D)}{dV_G}\right)^{-1} = \ln 10 \left(\frac{dN_S}{N_S dV_G}\right)^{-1} = \ln 10 \, \frac{eN_S}{C_{CH}}. \qquad (44)$$

Using Eqs. 22 and 34 this formula can be written down as follows

$$SS = \ln 10 \left(\frac{eN_S}{C_Q}\right)\left(1 + \frac{C_{it} + C_Q}{C_{ox}}\right) = \ln 10 \left(\frac{\varepsilon_F}{2e} + \frac{en_{res}}{C_{ox}}\frac{\varepsilon_a}{\varepsilon_F}\right)\left(1 + \frac{C_{it} + C_Q}{C_{ox}}\right). \tag{45}$$

Logarithmic swing should be minimized to obtain the maximum value of $I_{ON}/I_{OFF}$. In contrast to the Si-MOSFETs where SS = 70-100 mV/decade, such characteristics in gapless graphene cannot be achieved at any $C_{it}$ (even zero) and $\varepsilon_F$. This is a direct consequence of residual conductivity occurring in gapless graphene.

### *6.3. Low-frequency 1/f noise in graphene and traps*

The noise in the silicon MOSFET's is dominated at low frequencies (< 100 kHz) by flicker or 1/f noise. Flicker noise in graphene structures has been widely investigated experimentally recent years (see the review [41] and references therein). Influence of the charge traps in the $SiO_2$ was recognized as the primary source of 1/f noise in conventional silicon MOSFETs [42]. It is generally accepted that 1/f noise in FETs is described by the McWhorter model [43] which explains the characteristic spectral noise density by the carrier number fluctuations. It is assumed the low-frequency 1/f noise in FETs relates primarily to the exchange of carriers between the channel and defects located in the gate dielectric near the interface with the channel (slow interface traps or "border" traps [2]). Temporal fluctuations of occupancy of the traps with a broad range of characteristic recharging times result in temporal switching of the channel current (random telegraph noise [44]). Flicker noise with spectral power dependence $1/f^\alpha$ ($\alpha \sim 1$) can be considered as superposition of the random telegraph signals from the traps with the exponentially wide range of temporal constants arising due to tunneling [45]. According to the McWhorter model the correlation noise function

$$\frac{S_{I_D}(\Delta t)}{I_D^2} = \frac{\langle \delta Q_{ox}^2 \rangle}{Q_S^2} K(\Delta t), \qquad (46)$$

where $Q_S$ is total charge in the FET channel, $\langle \delta Q_{ox}^2 \rangle$ is dispersion of trapped oxide charge which is dependent on the energy level density of interface (border) traps near the Fermi energy in graphene. Time-dependent retarded response function characterizing temporal behavior of trapped charge can be deduced as [18]

$$K(t-t') = \frac{\lambda}{l}\left(E_1\left(\frac{t-t'}{\tau_{max}}\right) - E_1\left(\frac{t-t'}{\tau_{min}}\right)\right)\theta(t-t'), \qquad (47)$$

where $E_1(y)$ is the integral exponent function, the maximum and minimum times of tunneling recharging are connected as $\tau_{max} = \tau_{min}\exp(l/\lambda)$. Fourier-transform of the response function $K(\omega) = \int_0^\infty dt K(t)e^{+i\omega t}$ yields

$$K(\omega) = \frac{\lambda}{l}\frac{1}{\omega}\left\{\arctan\omega\tau_{max} - \arctan\omega\tau_{min} + \frac{i}{2}\ln\frac{1+(\omega\tau_{max})^2}{1+(\omega\tau_{min})^2}\right\}. \qquad (48)$$

Real part of the response function for the frequency range $\omega\tau_{max} \gg 1$ and $\omega\tau_{min} \to 0$ yields characteristic frequency dependence of flicker noise

$$\operatorname{Re} K(\omega) \cong \frac{\lambda}{l} \frac{1}{4f}. \tag{49}$$

This theory is generic for all types of FETs. In practice, with a large $Q_S$, the noise in graphene is typically higher than in conventional Si-MOSFETs, while a small $Q_S$ yields lesser noise in graphene FETs compared to the silicon FETs [41]. As expected the suspended graphene devices show a 6-12 times lower 1/f noise than those with an insulator substrate [46].

### *6.4. Ionizing radiation response of graphene field-effect devices*

One of the main possible application fields of graphene-based devices is spaceborne RF telecommunications systems. Therefore, it is important to discuss the impact of ionizing irradiation on performance of graphene field-effect transistors. Ionizing irradiation leads to an increase in the input gate capacitance and to a decrease in transconductance, field-effect mobility, and cutoff frequency. In general, these effects are similar to the effects in the silicon field-effect devices and represent the following [47]. When the gated graphene structure is exposed to ionizing radiation, the electron-hole pairs are created in the gate insulator. Under positive gate bias at room temperature, the radiation-induced electrons with relatively high mobility (~10 cm$^2$/Vs in the SiO$_2$) rapidly drift to the gate and easily leave the oxide, while holes with extremely low mobilities (~10$^{-5}$ cm$^2$/Vs) move slowly toward the graphene. A lesser part of these holes can create (directly or indirectly) the positively charge defects playing the roles of the fixed oxide charge and/or rechargeable interface traps with occupancy depending on the Fermi level position in graphene. Oxide charge accumulation leads to the shifts of the transfer I-V or C-V curve as a whole. The radiation-induced change of the oxide trapped charge $e\Delta N_{ot}$ which assumed to be located near the graphene (typically < 1-3 nm) can be characterized by a shift of

the CNP voltage

$$e\Delta N_{ot} = e\Delta N_t \left(\varepsilon_F = 0\right) = -C_{ox}\Delta V_{NP}. \tag{50}$$

Notice that the positive charge buildup is expressed in the negative shift $\Delta V_{NP}$ and vice-versa. It is commonly assumed that the hysteresis, often observed in graphene structures, originates from charge traps at the graphene/dielectric interface [48, 49, 50]. Typically the positive gate stress leads to electron capture onto the oxide traps, and the CNP shifts to the right. Delayed detrapping occurring at opposite direction of gate voltage sweeping has given rise to hysteresis in transfer curves in graphene devices as well as it often takes place in poor or irradiated MOS structures. Preliminary studies of the response of the graphene field-effect structures on $SiO_2$ to low-energy x-ray and gamma radiation exposure [51, 52, 53] show the effects basically determined by buildup of the positive oxide charge, interface traps, and degradation of carrier mobilities, like in conventional silicon MOS structures. Radiation-induced interface traps are capable to distort the shapes of transfer curves degrading transconductance, field-effect mobility and input gate capacitance. Distortion of the curve shapes was found to be typically asymmetric for electron and hole branches, and the CNP shifts towards negative gate voltages, because the trapped charge has positive sign. On the contrary, irradiation in oxygen atmosphere leads to significant positive shifts in $V_{NP}$ possibly due to oxygen adsorption [51]. It has been found that the minimum conductivity slightly increases after irradiation [53] likely due to occurrence of radiation induced electron-hole puddles at the CNP. Despite of the evidences about immediate breaking of the $sp_2$ bonds in graphene by the soft x-rays [54], the annealing at elevated temperatures recovers nearly all radiation-induced changes, implying these changes are not the effect of the lattice defects formation in graphene [53]. Suspended graphene transistors also degrade under X-ray exposure, but less than graphene-on-$SiO_2$ transistors [51]. Oxygen, hydrogen, and other reactive concentrations must be decreased before graphene FETs fabrication in order to achieve greater radiation tolerance [51, 52, 55]. Similar to conventional silicon MOS structures, radiation hardness of graphene transistors can be improved by thinning of gate and substrate dielectrics [56].Technological methods to minimize the effect of interface traps are reviewed in [10].

**References**

[1]     S.M. Sze, K.K. Ng, "Physics of Semiconductor Devices," 3rd edition, , John Wiley & Sons, ISBN 978-0-471-14323-9, New Jersey, USA. 2007.

[2 ]    D.M. Fleetwood, S.T. Pantelides, R.D. Schrimpf (Eds.) 2008, Defects in Microelectronic Materials and Devices, CRC Press Taylor & Francis Group, London - New York.

[3 ]    E. H. Hwang, S. Adam, S. Das Sarma, "Carrier Transport in Two-Dimensional Graphene Layers," Phys. Rev. Lett. 98, 186806 (2007).

[4 ]    S. Adam, E. H. Hwang, V. M. Galitski, and S. Das Sarma, Proc. Natl. Acad. Sci. U.S.A. 104, 18392 (2007).

[5 ]    E.H. Nicollian, J.R Brews, 1982, MOS (Metal Oxide Semiconductor) Physics and Technology, Bell Laboratories, Murray Hill, USA.

[6 ]    International Technology Roadmap For Semiconductors, 2009 Edition, Emerging Research Devices.

[7 ]    Byoung Hun Lee, Young Gon Lee, Uk Jin Jung, Yong Hun Kim, Hyeon Jun Hwang, Jin Ju Kim and Chang Goo Kang, " Issues with the electrical characterization of graphene devices," Carbon Letters Vol. 13, No. 1 (2012).

[8 ]    M. C. Lemme, T. J. Echtermeyer, M. Baus, and H. Kurz, "A Graphene Field-Effect Device," IEEE Electron Device Lett. 28, 282 (2007).

[9 ]    J. Zhu, R. Jhaveri, and J. C. S. Woo, "The effect of traps on the performance of graphene field-effect transistors," Applied Physics Letters 96, 193503 (2010).

[10]    G. Xu, Y. Zhang, X. Duan, A. A. Balandin, and Kang L. Wang "Variability Effects in Graphene: Challenges and Opportunities for Device Engineering and Applications," arXiv condmat 1302.3922.

[11]    G. I. Zebrev, "Graphene Field Effect Transistors: Diffusion-Drift Theory", a chapter in "Physics and Applications of Graphene – Theory", Ed. by S. Mikhailov, Intech, 2011.

[12 ]   S. Luryi, "Quantum Capacitance Devices," Applied Physics Letters, Vol. 52, 1988, pp. 501-503.

[13]    G.I. Zebrev, R.G. Useinov, "Simple model of current-voltage characteristics of a metal–insulator–semiconductor transistor", Fiz. Tekhn. Polupr. (Sov. Phys. Semiconductors), Vol. 24, No.5, 1990, pp. 777-781(1990).

[14 ]   S. Wolfram, Mathematica Book, Wolfram Media, ISBN 1–57955–022–3, USA, 2003.

[15]     J. Martin, N. Akerman, G. Ulbricht, T. Lohmann, J. H. Smet, K. von Klitzing, and A. Yacoby, ., "Observation of electron-hole puddles in graphene using a scanning single electron transistor," 2008, Nature Phys. 4, 144.

[16 ]    V.A. Gergel, R.A. Suris "Fluctuations of the surface potential in metal-insulator-conductor structures," JETP, 75, 191-203(1978).

[17]     G. Groeseneken, H. Maes et al., "Reliable Approach to Charge-Pumping Measurements in MOSTs," IEEE Trans. ED-31, no. 1, pp. 42-53, 1984.

[18 ]    V.V. Emelianov; G.I. Zebrev, V.N. Ulimov, R.G. Useinov, V.V. Belyakov, V.S. Pershenkov "Reversible positive charge annealing in MOS transistor during variety of electrical and thermal stresses," IEEE Trans. on. Nucl. Sci., 1996, No.3, Vol. 43, pp. 805-809.

[19]     H. Madan, M.J. Hollander et al., "Extraction of Near Interface Trap Density in Top Gated Graphene Transistor Using HF Current Voltage Characteristics," DOI: 10.1109/DRC.2012.6257022.

[20]     G.I. Zebrev, A.A. Tselykovskiy, D.K. Batmanova, E.V. Melnik, "Small-Signal Capacitance and Current Parameter Modeling in Large-Scale High-Frequency Graphene Field-Effect Transistors," IEEE Trans. On Electron Devices, Vol. 60, No. 6, 1799-1806(2013).

[21]     M.A. Ebrish, D. A. Deen, S. J. Koester, "Border Trap Characterization in Metal-Oxide-Graphene Capacitors with HfO2 Dielectrics," 2013.

[22]     Chen, 2013 Xiaolong Chen, Lin Wang, Wei Li, Yang Wang, Zefei Wu, Mingwei Zhang, Yu Han, Yuheng He, and Ning Wang, "Electron-electron interactions in monolayer graphene quantum capacitors," Nano Research, DOI 10.1007/s12274-013-0338-2, 2013.

[23]     M.A. Ebrish, S. J. Koester, "Dielectric Thickness Dependence of Quantum Capacitance in Graphene Varactors with Local Metal Back Gates," DOI 10.1109/DRC.2012.6256974, Device Research Conference, pp. 105-106, 2012.

[24]     S.-J. Han, D. Reddy, G. D. Carpenter, A. D. Franklin, and K. A. Jenkins, "Current Saturation in Submicrometer Graphene Transistors with Thin Gate Dielectric: Experiment, Simulation, and Theory," ACS Nano, 2012 Jun 26; 6(6):5220-6. doi: 10.1021/nn300978c.

[25 ]    T. Fang, A. Konar, H. Xing, and D. Jena, "Carrier statistics and quantum capacitance of graphene sheets and ribbons," Appl. Phys. Let. 91, 092109 (2007).

[26]     G.I. Zebrev, "Graphene nanoelectronics: electrostatics and kinetics", Proceedings SPIE, 2008, Vol. 7025. – P. 70250M - 70250M-9, based on report to ICMNE-2007, October, 2007, Russia.

[27]     D. C. Elias, R. V. Gorbachev, A. S. Mayorov, S. V. Morozov, A. A. Zhukov, P. Blake, L. A. Ponomarenko, I. V. Grigorieva, K. S. Novoselov, F. Guinea, A.K. Geim, "Dirac cones reshaped

by interaction effects in suspended graphene," Nat Phys 7(9): 701-704 (2011).

[28] F. de Juan, A. G. Grushi and M. A. H. Vozmediano, "Renormalization of Coulomb interaction in graphene: Determining observable quantities," Physical Review B 82, 125409 (2010).

[29] G.L. Yu, R. Jalili, B. Belle, A.S. Mayorov, P. Blake, F. Shedin, S. V. Morozov, L. A. Ponomarenko, F. Chiappini, S. Wideman, U. Zeitler, M.I. Katsnelson, A.K. Geim, K.S. Novoselov, and D. Elias, "Interaction Phenomena in Graphene Seen Through Quantum Capacitance," www.pnas.org/cgi/doi/pnas/1300599110 (2013).

[30] G.I. Zebrev, E.V. Melnik, "Using Capacitance Methods for Interface Trap Level Density Extraction in Graphene Field-Effect Devices," Proc. of 28th International Conference on Microelectronics MIEL 2012, 335-338(2012), doi:10.1109/MIEL.2012.6222868.

[31] L.A. Ponomarenko, R. Yang, R.V. Gorbachev, P. Blake, A. S. Mayorov, K. S. Novoselov, M. I. Katsnelson, and A. K. Geim, "Density of States and Zero Landau Level Probed through Capacitance of Graphene," Phys. Rev. Lett. 105, 136801 (2010).

[32] Z. Chen, and J. Appenzeller, "Mobility extraction and quantum capacitance impact in high performance graphene field-effect transistor devices," in IEEE International Electron Devices Meeting 2008, Technical Digest, p. 509-512 (2008).

[33] A. A. Kozikov, A. K. Savchenko, B. N. Narozhny, A. V. Shytov "Electron-electron interactions in the conductivity of graphene," Phys. Rev. B 82, 075424 82, 075424 (2010).

[34] S. Russo, M.F. Craciun, M. Yamamoto, A.F. Morpurgo, S. Tarucha, "Contact resistance in graphene-based devices", Physica E, Volume 42, Issue 4, p. 677-679, 2010.

[35] Y.G. Lee, Y.J.Kim, C.G. Kang, C. Cho, S. Lee, H.J. Hwang, U.Jung, B.H. Lee, "Influence of intrinsic factors on accuracy of mobility extraction in graphene field effect transistors," Appl.Phys. Lett. 102, 093121(2013).

[36] M. Latkiofi, B. Krauss, T. Lohmann, U. Zschieschang, H. Klauk, K. v. Klitzing, J.H. Smet, "Graphene on a Hydrophobic Substrate: Doping Reduction and Hysteresis Suppression under Ambient Conditions," Nano Lett. 10, 1149 (2010).

[37] D. Nezich, A. Reina, and J. Kong "Electrical characterization of graphene synthesized by CVD using Ni substrate," Nanotechnology 23, 015701 (2012), doi:10.1088/0957-4484/23/1/015701.

[38] F. Schwierz, "Graphene Transistors," Nature Nanotechnology, 30 May 2010 | doi: 10.1038/nnano.2010.89

[39] G.I. Zebrev, "Electrostatics and diffusion-drift transport in graphene field effect transistors," Proc. of 26th International Conference on Microelectronics MIEL 2008, 159-162(2008), doi:10.1109/ICMEL.2008.4559247.

[40]   T. Ando, A. Fowler, F. Stern, "Electronic properties of two-dimensional systems," Rev. Mod. Phys. Vol. 54, No.2, 1982, pp.437-462.

[41]   A. A. Balandin, "Review of the low-frequency 1/f noise in graphene devices, arXiv cond-mat 1307.4797.

[42]   J.H. Scofield, D.M. Fleetwood, "Physical basis for nondestructive tests of MOS radiation hardness," IEEE Trans on Nucl. Sci. Vol. 38, 1567-77 (1991).

[43]   A.L. McWhorter, Semiconductor Surface Physics, Ed. By R.H. Kingston, p. 207, University of Pennsylvania Press (1957).

[44]   K. K. Hung, P. K. Ko, C. Hu and Y. C. Cheng, "Random telegraph noise of deep-submicrometer MOSFETs," Electron Device Letters, IEEE, vol. 11, pp. 90-92, 1990.

[45]   M. J. Uren, D. J. Day and M. J. Kirton, "1/f and random telegraph noise in silicon metal oxide semiconductor field effect transistors," Applied Physics Letters, vol. 47, no. 11, 1195-1197, 1985.

[46]   Z. Cheng, Q. Li, Z. Li, Q. Zhou and Y. Fang, "Suspended Graphene Sensors with Improved Signal and Reduced Noise," Nano Letters, vol. 10, pp.1864-1868, 2010.

[47]   T.R. Oldham, F. B. McLean, "Total ionizing dose effects in MOS oxides and devices," IEEE Transactions on Nuclear Science, Vol. 50, no. 3, pp. 483–499, 2003.

[48]   H. Wang, Y. Wu, C. Cong, J. Shang, and T. Yu, "Hysteresis of electronic transport in graphene transistors," ACS Nano, vol. 4, no. 12, pp. 7221–7228, 2010.

[49]   Y. G. Lee, C. G. Kang, U. J. Jung, J. J. Kim, H. J. Hwang, H. J. Chung, S. Seo, R. Choi, and B. H. Lee, "Fast transient charging at the graphene/SiO2 interface causing hysteretic device characteristics," Applied Physics Letters, vol. 98, p. 183508, 2010.

[50]   C. X. Zhang, E. X. Zhang, D. M. Fleetwood, M. L. Alles, R. D. Schrimpf, E. B. Song, S. M. Kim, K. Galatsis, K. L. Wang, "Electrical Stress and Total Ionizing Dose Effects on Graphene-Based Non-Volatile Memory Devices," IEEE Transactions on Nuclear Science, Vol. 59, No. 6, Dec 2012.

[51]   E. X. Zhang, A. Newaz, S. Bhandaru, B. Wang, C. X. Zhang, D. M. Fleetwood, M. L. Alles, S. T. Pantelides, S. M. Weiss, Kirill I. Bolotin, R. A. Reed, R. A. Weller, "Low-Energy X-ray and Ozone-Exposure Induced Defect Formation in Graphene Materials and Devices," IEEE Transactions on Nuclear Science, Vol. 58, No. 6, Dec 2011.

[52]   E. X. Zhang, A. K. M. Newaz, B. Wang, C. X. Zhang, D. M. Fleetwood, K. I. Bolotin, R. D. Schrimpf, S. T. Pantelides, M. L.Alles, "Ozone-exposure and annealing effects on graphene-on-SiO2 transistors," Appl. Phys. Lett. 101, 121601 (2012); doi: 10.1063/1.4753817.

[53] C.D. Cress, J.G. Champlain, I.S. Esqueda, J.T. Robinson, A.L. Friedman, J.J. McMorrow, "Total Ionizing Dose Induced Charge Carrier Scattering in Graphene Devices," IEEE Transactions on Nuclear Science, Vol. 59, No. 6, Dec 2012.

[54] S. Y. Zhou, C.O. Grit, A. Sholl et al., "Instability of two-dimensional graphene", Phys. Rev. B 80, 121409, (2009).

[55] Y. S. Puzyrev, B. Wang, E. X. Zhang, C. X. Zhang, A. K. M. Newaz, K. Bolotin, D. M. Fleetwood, R. D. Schrimpf, S. T. Pantelides, "Surface reactions and defect formation in irradiated graphene devices," IEEE Transactions on Nuclear Science, 59, 3039 (2012).

[56] P.E. Dodd, M.R. Shaneyfelt, J.R. Schwank, J.A. Felix, "Current and Future Challenges in Radiation Effects on CMOS Electronics," IEEE Transactions on Nuclear Science, Vol. 57, no. 4, pp. 1747-63, Aug. 2010.